# Graphite in Science and Nuclear Technology

This review is devoted to the application of graphite and graphite composites in the science and technology. Structure and electrical properties, technological aspects of producing of high-strength artificial graphite and dynamics of its destruction are considered.

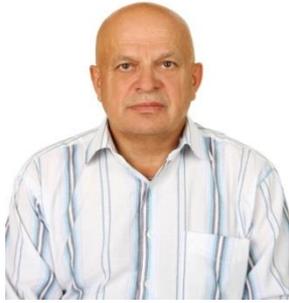

This type of graphite are traditionally used in the nuclear industry, so author concentrates on the actual problems of application and testing of graphite materials in the modern science and technology.

Translated from chapters 1 of the monograph (by *Zhmurikov E.I., Bubnenkov I.A., Pokrovsky A.S. et al.* Graphite in Science and Nuclear Technique// eprint arXiv:1307.1869, 07/2013 (BC 2013arXiv1307.1869Z).

Author: *Evgenij I. Zhmurikov*

Address: *630090 Novosibirsk, Russia*

E-mail: *evg.zhmurikov@gmail.com*

**Introduction**

The graphite has the unique set of qualities that make it as an ideal material for problems in the nuclear physics and nuclear power engineering. Natural carbon is a mixture of two stable isotopes: $^{12}$C (98,892%) and $^{13}$C (1,108 %). Only the $^{14}$C isotope with a half–life of 5730 years is a long–lived isotope from four radioactive isotopes ($^{10}$C, $^{11}$C, $^{14}$C and $^{15}$C) [1]. The $^{14}$C isotope is pure low energy ß–emitter with a maximum energy of particles equal 156 keV and $^{14}$C refers to global radionuclides. However, the radiation threshold for formation of this radioisotope is quite high, so this radioisotope is mainly produced in nuclear explosions or by the interaction of secondary neutrons with cosmic radiation. It emerges by the reaction of nitrogen nuclei $^{14}$N (n, p) => $^{14}$C. The role of other reactions in the formation of the isotope $^{14}$C is extremely small.

Other features of the graphite for nuclear physics associated with a small cross section $\sigma$ of photo–nuclear reactions for carbon in the giant resonance region. It is connected with the excitation by the γ–quanta of own oscillations of protons relative to neutrons (dipole oscillations). The nucleons can leave the nucleus, not only in the process of dipole oscillations, but also after they damping.

Thus, the secondary radiation is relatively small for pure graphite under irradiation even fairly high–energy (up to 50 MeV) proton beam. This is due to the small cross section of secondary neutrons in the reaction with carbon nuclei – less than 4.5 microbarns for high–purity graphite [2]. Most of the neutron collisions with nuclei of carbon occur through a mechanism of elastic scattering; the latter led to the effective using of graphite as a moderator or neutron absorber.

In particular, graphite as an efficiency moderator comes after beryllium and heavy water for nuclear reactors that are operating on the enriched uranium. In this case high purity graphite is used, wherein the total content of impurities does not exceed $1 \times 10^{-3}$ %. More pure graphite was created for using in semiconductor technology with impurity content at most $1 \times 10^{-6}$ %.

Graphite is a good construction material, and remains solid up to sublimation temperature about $4000^{0}$C. At the same time graphite has a relatively low density and not only sufficiently strong, but also plastic, easily processed mechanically, and have a low vapor pressure even under vacuum at elevated temperature. In addition, graphite has high thermal conductivity, heat capacity and not very high electro conductivity. Strength and plasticity of graphite noticeably increases with temperature up to ~ 2500$^{o}$C [2]. Moreover, graphite is resistant to both thermal shock and a high temperature gradient due to the high porosity and able to deliver of excess heat by the reemission in infrared and optical range. Graphite is irreplaceable in scientific and practical applications due to excellent corrosion and chemical resistance in combination with anti-friction properties.

Graphite is not oxidized up to temperature of $400^{\circ}C$ in air, and up to $500^{\circ}C$ in carbon dioxide, but graphite items have to be used in a protective atmosphere or in vacuum at higher temperatures.

At the same time, graphite as a construction material has been studied not quite enough. In particular, there are not clear reasons for the strong scattering of the physical-mechanical and thermal properties of graphite for various grades of graphite and even within the same brand. Properties of the graphite in difficult conditions of enhanced radiation and high heating have not been sufficiently studied, and it is not quite clear the strong anisotropy for well graphitized materials. The solidity of the graphite can be considerably varies according to the method of its manufacture. Therefore graphite with the equal density due to their structure differences may be of different durability. The general rule is that the more finely structured graphite composite has the more greater solidity and the more long lifetime.

It should be noted that as far as possible the author have tried to follow of the common scientific terminology, that is offered in [3] or [4]. However, it is important to understand that for some reason is not always possible to come after the strict terms. In particular, in this book as in other, the word "graphite" is often used broadly for shortness, instead of strict definition like the "..artificial graphite obtained by the technology of the electrode using the carbon powder as a filler".

In addition, the glossary [4] is defined the word "composite" as a material or a material reinforced by a fibrous filler. However, this definition does not contradict, we guess, its using for carbon materials with dispersion carbon filler, too. Therefore, the word "composite" have been used for material that is based on the fine powder of isotope $^{13}C$, although the term in this case is not a strict.

## 1. The main concepts about the graphite structure and properties. Dynamics of the destruction, strength and durability of the graphite composite

Graphite is a thermodynamically stable allotropic modification of carbon – an element of the fourth group of the Mendeleev periodic table. Carbon has number 6 in this table, it is located in the 2–nd period of the main subgroup and the atomic mass of the natural mixture of isotopes is 12,0107 g/mol [5].

It should be noted that carbon exists in several allotropic modifications in various isomorphic states with very diverse physical properties [6–9]. The capability to form chemical bonds of various types determined by variety of carbon allotropic modifications. Electronic orbitals of carbon atoms can be different geometries, depending on the degree of hybridization. There are three basic geometry of the carbon atom: tetrahedral, trigonal and digonal. Diamond and lonsdalite correspond to tetrahedral geometry of carbon atoms. Carbon has the same hybridization, for example, in methane and other hydrocarbons.

Trigonal modification is formed by mixing one *s*– and two *p*–electron orbitals (*sp²*–hybridization). Atom of the carbon has three equivalent $\sigma_{x,y}$–bonds, arranged in one plane at the angle of 120° to each other. In this case $p_z$–orbital is not involved in the hybridization, located perpendicularly to the plane of the σ–bonds, and used to form σ–connection with other atoms. This is namely geometry of carbon is typical for the graphite.

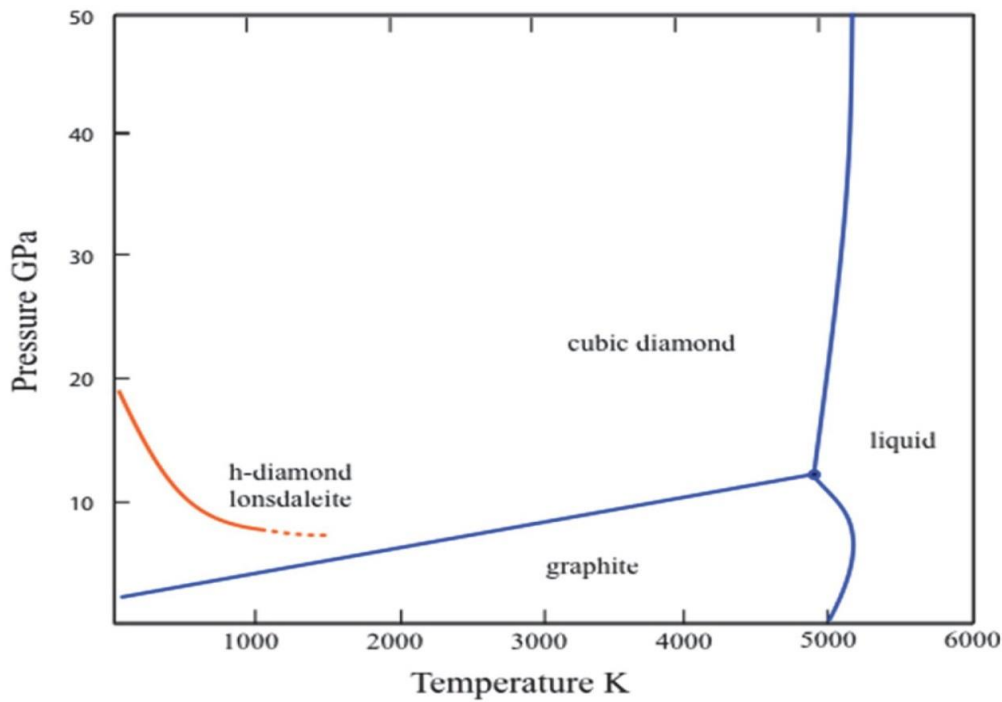

**Fig**. **1.1**. The phase diagram by *Bundy et al.*, 1996 [15, 16]

Carbon with digonal atom geometry forms a special allotropic modification – carbyne. However, until now length of the carbyne ligaments without bends and interchain connection is less than a hundred atoms [10]. At ordinary temperatures, carbon is chemically inert, but can connected with a lot of elements at high temperatures, showing strong restorative properties. The Bundy's phase diagram is shown in *fig*. 1.1, various allotropic modification of carbon and its isomorphism are shown in *fig*. 1.2.

It should be noted that this phase diagram is not considered as undoubted; in particular, until now the question about of the melting point of graphitet [11, 12], because, according [12], there is a "...an significant divergence between the data connected with the melting point of graphite". Nevertheless, most of the authors are agree with the phase diagram Bundy, the question is, as a rule, only about a little revisions diagram in varying regions of the pressures and temperatures [13, 14].

## 1.1. **Structural and electrophysical properties of graphite**

### 1.1.1. The crystal lattice of carbon

In graphite, each carbon atom covalently linked to three other carbon atoms surrounding it. There are hexagonal and rhombohedral modifications of graphite [5], which differ by packing of layers (*fig*.1.3).In hexagonal graphite half of the atoms of each layer is located above and below the center of the hexagon (ABABABA motive …), while in rhombohedral one every fourth layer repeats the first.

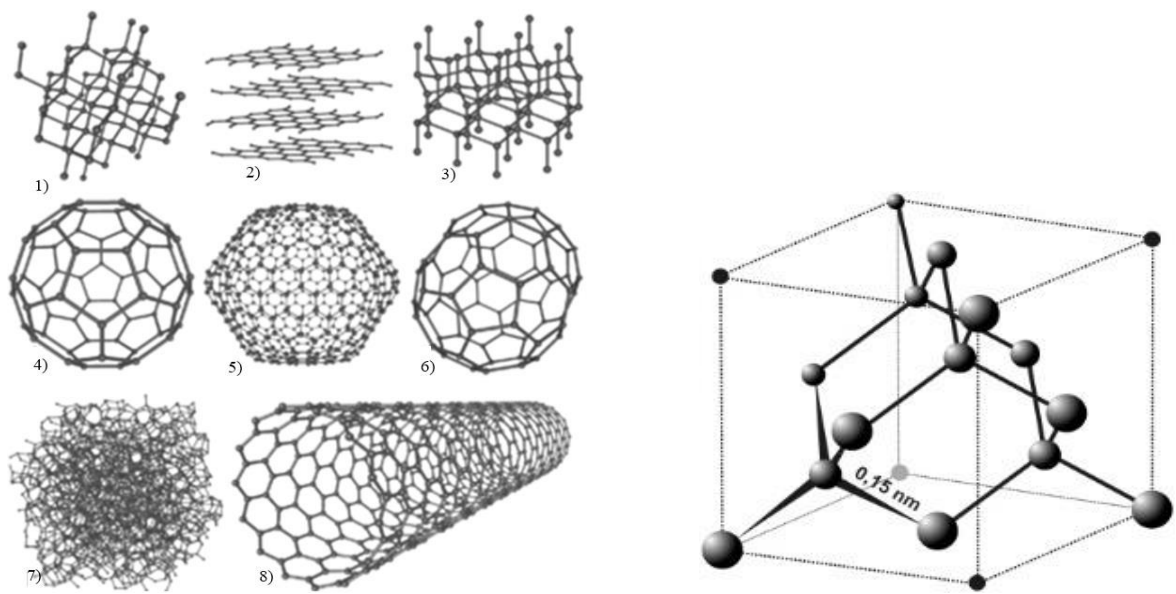

**Fig**. **1.2**, *left*) **–** Scheme of the structure of carbon modifications:*left*
1) diamond; 2)graphite; 3) lonsdalite; 4) Fullerene – C60; 5) fullerene – C540; 6) fullerene – C70; 7) amorphous carbon; 8) carbon nanotube. *right* – elementary cell of the diamond.

The rhombohedral graphite conveniently presented in hexagonal axes to show its layered structure. Rhombohedral graphite is not observed in its pure form, because its metastable phase. However, the rhombohedral phase content can reach 30 % in natural graphite.

The hexagonal graphite lattice belongs to the space group C6 / mmc– $D^4_{6h}$ with four atoms per elementary cell. Parameter **a** of the hexagonal cell is 2,46Å, parameter **c** = 6,71Å, theoretical density of this crystal is 2,267 g/cm$^3$. In each plane of the carbon atoms form a network of regular hexagons with the interatomic distance 1,42Å. Links within the layers are covalent and forms trigonal hybrids ($2s$, $2p_x$, $2p_y$) [5]. The atomic connection in layers is carried out by the covalent σ–bonds, and links between layers are providing by relatively weak π connection. There is sometimes written the connection between the layers is carried out by van der Waals forces, with some contribution of polarization forces.

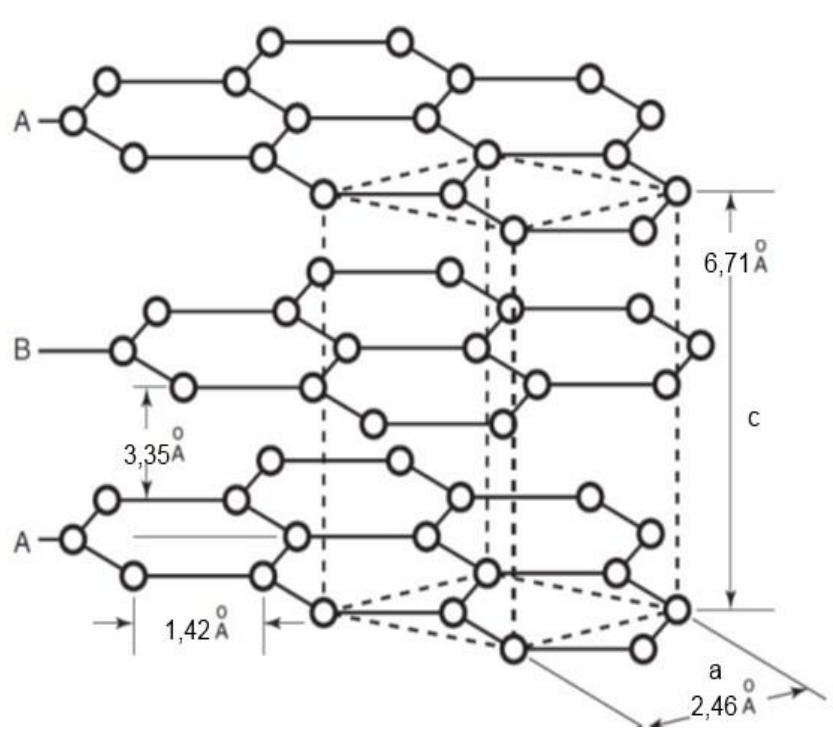

**Fig**. **1.3**. The crystal lattice of hexagonal graphite [5].The crystal lattice of a two–layer with alternation of layers in the c–direction: ...ABA–ABA...; Z=4; the coordination number is – 3; the unit cell parameters: **a** = 2,46 Å, **c** =6,71 Å; length of σ–bond is 1,42 Å, π–bond – 3,35 Å; energy of σ–bond= 418 – 461 kJ/mol; π bond – 16,8 – 41,9 kJ/mol; calculated roentgen graphic density – 2,267 g/cm$^3$.

1.1.2. X–ray diffraction studies of polycrystalline graphite based on the fact that under the Wulff–Bragg condition peaks of the diffraction pattern caused by the reflection of X-rays from a system of parallel crystallographic planes. The path difference is equal to the whole number of wavelengths, when the rays reflected from different planes of this system:

$$2d \sin \theta = m\lambda, \qquad (1.1)$$

where $d$ — interlayer spacing, $\theta$ — slipping angle, i.e. the angle between the reflecting plane and the fall beam, $\lambda$ is the x-ray wavelength , and $m$ is the so-called order of reflection, i.e. positive integer number. Investigation of the polycrystalline materials structure using X-ray diffraction method based on Debye-Scherrer method [17]. Metals, alloys, crystalline powders are consisted of a plurality of small single crystals and so to study its structure one can use monochromatic radiation.

In this method, a narrow parallel beam of monochromatic X-rays falling on the polycrystalline sample, bouncing off from the crystals and gives a series of coaxial, i.e. there are diffraction cones with common axis. The axis of the cone is the direction of the primary beam of X-rays, they tops are inside of the object under study, and the angles of the reflection are determined according to the Wulff-Bragg equation.

The angle of the cone is equal to quadruple the angle of reflection $\theta$, and by measuring the angles of diffraction cones one can to determine the interplanar distances $d$. In some cases, these data combined with the measurement of the intensity of each x-ray diffraction cone is sufficient for the complete determination of the crystal lattice structure.

Roentgenogram (powder patterns) of polycrystals consists of several concentric rings, each of which merge reflection from system of planes variously oriented single crystals. Powder patterns for the miscellaneous materials are individual and are widely used for the identification of compounds and mixtures. X-ray analysis of polycrystalline allows determining the phase composition of samples, establishing the size and preferred orientation (texturing) grains in the material, and allows monitoring the internal stresses, too. The lower will be the ordering of the atomic structure of the material, the more vague, diffuse will be scattered X-rays.

The number $n$ of point reflections of Debye ring corresponds to the number $N$ of large crystallites involved in the reflection of X-rays:

$$N = 2n/\alpha \cos\vartheta, \qquad (1.2)$$

where $\alpha$ is constant (hardware option) and $\vartheta$ — the Bragg angle.

Small crystallite sizes $L$ can be estimated from the width of the diffraction maxima using Selyakova – Scherrer's formula [18, 19]:

$$L = L_0 \cdot \lambda / (\beta \cdot \cos \vartheta), \qquad (1.3)$$

where   $L_0$ - constant, depending on the shape of the particles;

   β - line width at the half maximum;

   λ — the wavelength of the x–ray reflection;

   ϑ - the Bragg's diffraction angle.

The shape of the crystallites of carbon materials can be considering as disk, and then one can determine the diameter of the crystallites *La* from the width of the line **hko**, the crystallite height *Lc* is determined from the width of the line **00l**. The value of microdistortions for crystal lattice can be estimated using the ratio between the diffraction peaks with Miller indices 002 and 004. However, as the size of the crystallites and microdistortions value can be obtained only with a certain reliability in the study of the diffraction lines by methods of Fourier analysis [20].

### 1.1.3. Structural defects in graphite. The types of defects

Defects in graphite [21, 22] can be divided into two types: relating to violations between layers and defects of layers. The first group includes stacking faults of layers, these defects are characterized by violation of the order of the parallel layers package of hexagonal lattice. The carbon usually called as *turbostratic,* if it consisting of more or less perfect hexagonal grid, but with violation of the order in the sequence of layers of packaging. Graphite layers are displaced relative to each other randomly (with random vector displacement of one layer to another) in such type of a structure. Bonds defects are the second type of violation of the structure in graphite. This are vacancies and their groups, impurity atoms implanted in a hexagonal layer, defects of isomeric bonds, when part of the atoms are $sp^3$ – hybridization, edge defects, etc.

One can see the main types of the second type of defects. There are:

*Edge defects* when C–C bond can be formed, for example, if one macromolecule is not in the plane of its nearest neighbors. Foreign atoms often interact with edge defects, such as H or group like – OH, = O, –O– [22]. Furthermore, the carbon atoms at the edge defects are form weak bonds with neighboring atoms, which will be disintegrate at heating.

*"Splitting" defects* appear, when the voids or gaps formed by the destruction of hexagonal lattice of carbon atoms. Screw dislocations or curved hexagonal layers or lattice can occur by means of "splitting" defects .

*Defects of isomeric bonds* of lattice configurations, because expected that carbon atoms retain in $sp^2$–hybridization. However, some of the atoms can be connected with carbon atoms by means of sp³ hybridization. Firstly it is take a place in very small crystallites that usually have a lot of the edge defects.

*Defects of "twinning"*, when twinning appears on the line by rings consisting of four or eight carbon atoms. Should be noted that there are two types of twinning: in the first case it is in the form of accretion, with axis is parallel to the axis "$c$" of the graphite lattice. It is called the basic twinning, there are crystalline formations of two or more pieces, the same by the composition and structure, but not identical in shape and size. They are regularly spaced relative to each other in composite. The regularity is that one part of the lattice is aligned with the other rotating in a twinning axis. The second type of twinning is extrabasis, it appears due to the reflection in the twin plane or by combining rotation and reflection. Extrabasis twinning requires a plane of symmetry (mirror plane called the twinning plane). It looks like bending of graphite planes at a certain angle.

There are only two corner of true twinning (48°18' and 35°12') for hexagonal graphite structure with a sequence ABAB of layers packaging. In other cases, the symmetry laws are violated, and sequence of layers on each side of the interface is not stored. Hence, the inclined crystal boundaries, symmetrical and especially asymmetrical must be accompanied of many breaks links and a well–developed system of edge dislocations.

*Chemical defects* occur as inclusions of foreign atoms in the carbon lattice. In addition, there are displacements of atoms from their normal positions in the lattice. Thus, in the graphite lattice can be the atoms O, S, Se, As, N, P and others. One can expect that the atoms in the periodic table of the elements left of carbon will serve as the electron acceptors and lying to the rights side from carbon will be donors. In particular, the boron atoms are the acceptors, leading to a positive temperature coefficient of conductivity.

*Layer defects* include vacancies and its groups; it can be Frenkel or Schottky defects, and edge or screw dislocations. These defects appear, for example, by cause of neutron irradiation. Carbon atoms under the irradiation are displaced from their normal positions, forming the closing links between the carbon macromolecules. In highly irradiated graphite crystals can be extended interlayer spacing, this disturbance is removed by annealing at 1500°C.

### 1.1.4. Electronic structure, electrical and thermal properties of polycrystalline graphite

Each carbon atom in the graphite has four valence electrons, three of which form strong covalent σ–bonds in the hexagonal graphene layer. The fourth electron remains in the π–state, providing a weak covalent bond between neighboring layers of graphene.

According to numerous calculations and X-ray studies [18] the three bonding and three antibonding molecular orbitals (MO) [23, 24] are formed the filled and empty σ-zones, which are separated by an energy interval of ~ 5 eV about. This gap includes the Fermi level and π-zone, too. The reciprocal crystallographic lattice is formed from the rules [25] one can get the first Brillouin zone of graphite in the form of a right prism with a hexagonal base, if in the first approximation neglect the interaction between the graphene layers [26].

The first Brillouin zone of graphite is shown on the fig. 1.4 (a), here $k_x$, $k_y$, $k_z$ are the projections of the wave vectors. On the fig. 1.4 (b) is shown a two-dimensional Brillouin zone. This zone is built in the reciprocal space (in the space of quasi-impulses) upon condition $k_z$=0. Model wave functions used in the calculations have not any component in the $c$-direction of (or in the $z$-direction in the reciprocal space). These functions are commonly indicated as σ, and the wave function with only $z$-component as π, respectively. Appropriate zone are called as σ – and π – zone.

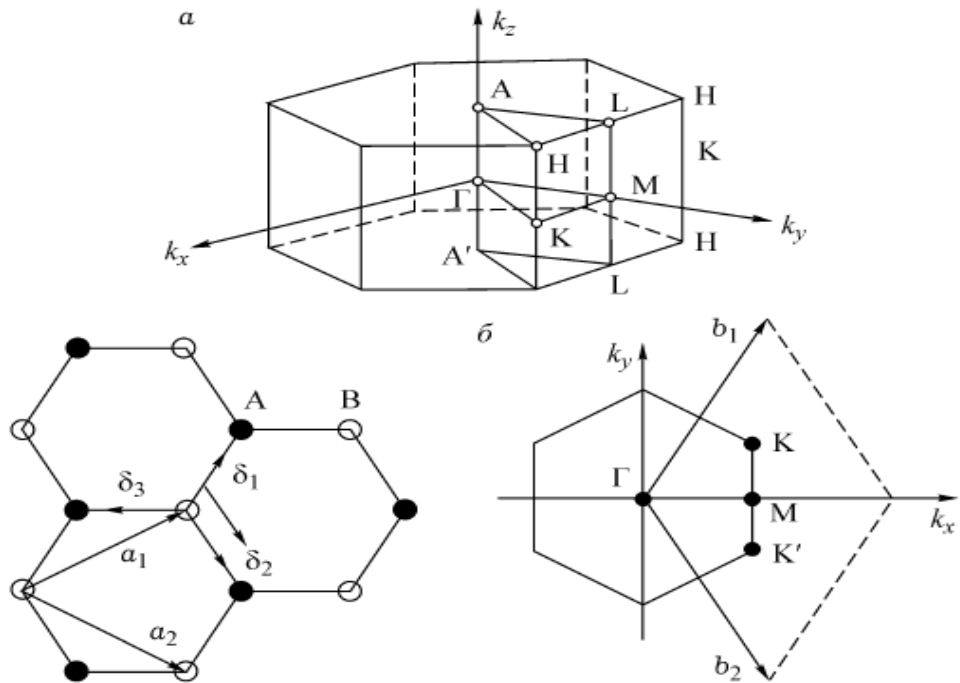

**Fig**. **1.4**   a) The primitive cell of the first Brillouin zone for the hexagonal graphite. b) The hexagonal graphene lattice with translation vectors $a_1$, $a_2$ and the first Brillouin zone of the reciprocal lattice with vectors $b_1$, and $b_2$ [29].

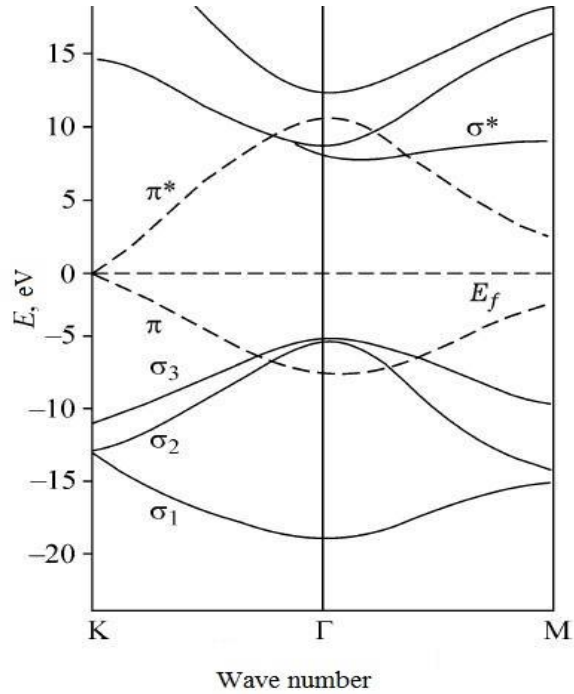

**Fig**. **1**.**5**. The results of two–dimensional calculation of the dispersion of energy along of two characteristic lines of the Brillouin zone [26].

Strictly speaking, a separation to σ– and π – areas is valid only for highly symmetric points and directions subzone ΓALHKM, and it is only 1/24 of the entire graphite Brillouin zone. For two-dimensional calculations of energy dispersion in graphite ⟨we can neglect the interaction between the layers as a first approximation⟩ are generally used one–electron and the adiabatic approach. An example of calculating the energy dispersion toward the wave vector is shown on the fig.1.5.

The band structure of graphite was first calculated in [27] in the strong coupling approximation. If we neglect the interaction between the graphene layers, then, as has been said, the Brillouin zone is reduced to hex ⟨fig. 1.6⟩, and the primitive cell contains two atoms. The dependence of the electron state density toward energy according to [28] is shown in *fig. 1.7*.

Near the contact points of the valence band and the conduction band ⟨*K* and *K*⟩ the dispersion law for charge carriers ⟨electrons⟩ in graphene has the form as shown on fig. 1.8.

According to [29]:

$$E = \hbar v_F k, \qquad (1.4),$$

where: $v_F$ – Fermi velocity ⟨$v_F \sim 10^6$ m/sec⟩;

$k$ – modulus of the wave vector in the two–dimensional space, that counted from $K$ and $K'$ Dirac points with the components $k_x$ and $k_y$;

$\hbar$– Planck's constant.

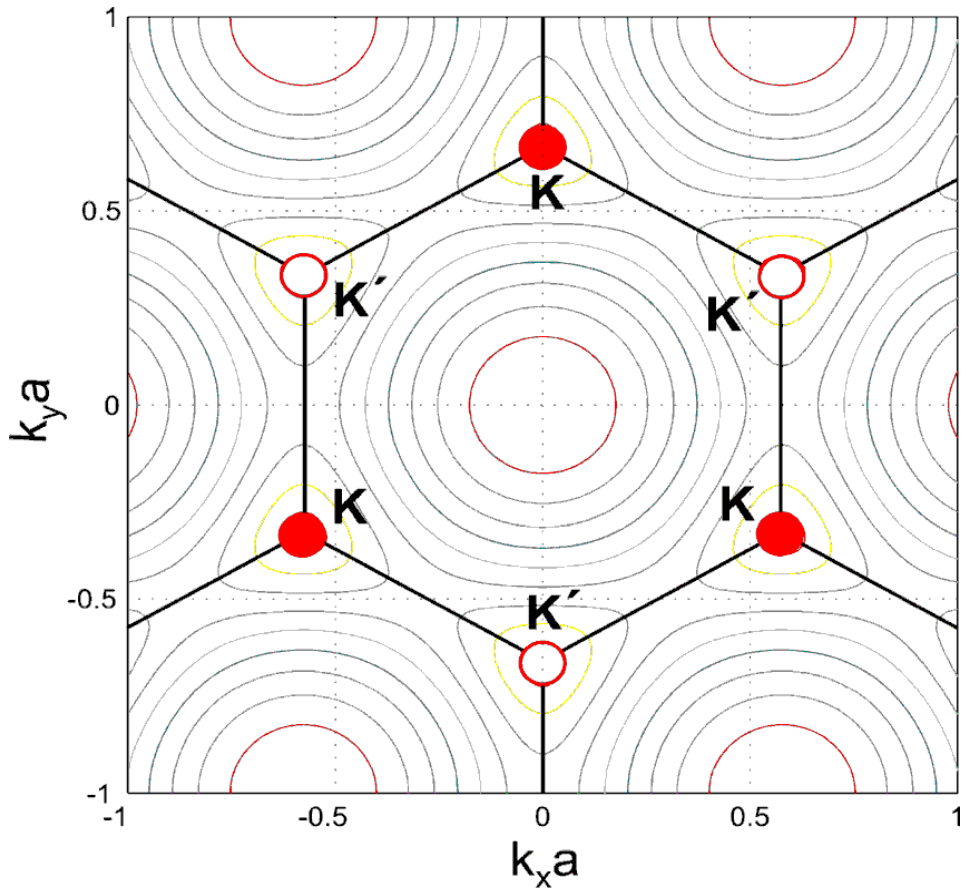

**Fig**. **1.6**. Isolines of constant energy for the graphene layer. Bold black hexagon is the first Brillouin zone. Inner circle correspond to the edges of the first Brillouin zone, where the carrier dispersion law is linear. $K$ and K' denote the two valleys in $k$–space with a non–equivalent wave vectors.

The linear dispersion law leads to a linear dependence of density of states upon the energy, unlike the usual two–dimensional systems with a parabolic dispersion law, where the density of states does not depend on energy. The density of electron states per unit of area is [29]

$$\nu(E) = g_s g_v \, |E| \, /2\pi \nu \vartheta_F^2 \qquad (1.5),$$

where $g_s$ and $g_v$— double spin and three times valley degeneracy, respectively; the module of energy is required to describe the electrons and holes by an one formula. We can see therefore that at zero the energy density of states is zero too, there are no carriers (at zero temperature).

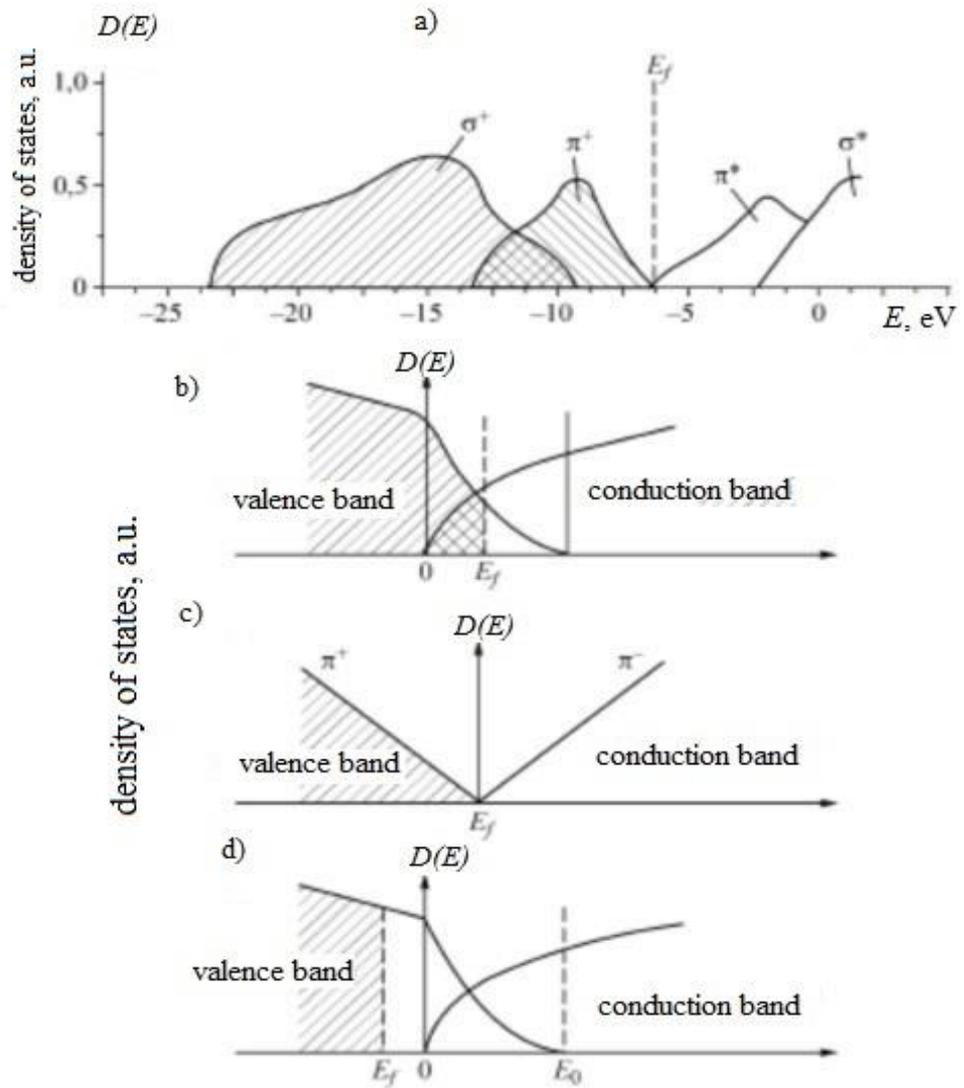

**Fig**. **1.7**. a) The dependence of the electron density of states via the energy in the crystalline graphite. A zero energy counted from: a) – the level of vacuum; b) – from the bottom of the antibonding $\pi$ – zone; c) the density of states for the graphene plane; d) for boron doped crystalline graphite [28].

The band structure of three–dimensional model of graphite differs from the two–dimensional only in the vicinity of contact between the valence band and the conduction band due to the small interaction between the layers. However, this small interaction qualitatively changes the Fermi surface. Most of the electronic properties of graphite, in particular, the electrical conductivity can be understood only by a three–dimensional model. The interlayer interactions account leads to a dependence of the $\pi$–electrons energy from the wave vector $k_z$ along the edge HKH (fig. 1.4). The mostly important result of the interlayer interactions is the overlap of the valence and conduction band on the value of ~ 0,03 ÷ 0,04 eV, as well as the distortion of the linear dispersion law in the energy range from 0 ~ 0.5 eV near the edge HKH zone (fig. 1.4). Overlapping zones leads to equal concentration both of electrons and holes ~3·10$^{18}$ cm$^{-3}$ at 0 K, i.e. monocrystal of graphite is a typical semimetal [18].

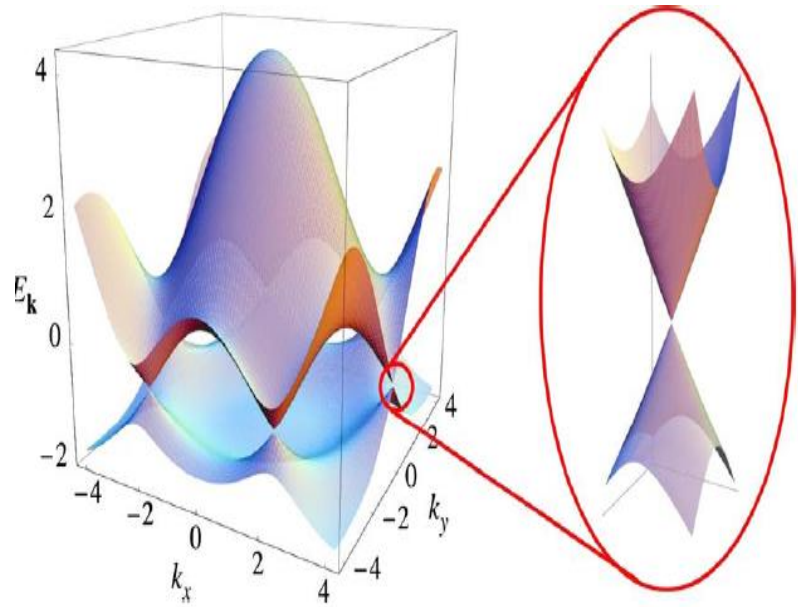

**Fig**. **1**.**8**. The electron density of states in the "honeycomb" of the graphite crystallographic lattice. The energy spectrum is presented in arbitrary units, to right with an increase is shown the contact between valence and conduction band [29].

Investigations by means *de Haas–van Alphen and Shubnikov–de Haas* oscillation phenomena allowed to specify the type of the Fermi surface and the effective masses of the carriers in a monocrystalline of graphite. The Fermi surface in first approximations is an ellipsoid of rotation with an axial ratio of 12.6: 1 for holes and 11: 1 for the electrons. In this case, the major axes coincide with the direction of the crystallographic *c*-axis of graphite. The electrons and hole effective masses are equal to 0,06m$_0$ and 0,04m$_0$ for carrier movement along the graphene layer. As far as the electronic properties of monocrystalline graphite has a significant anisotropy, so it will be 14m$_0$ and 5,7m$_0$ to move perpendicular to the layer for the electrons and hole, respectively [30].

The anisotropy of crystalline graphite significantly affects to the thermal conductivity of graphite, where appear its semiconductor properties [31]. Graphite materials have characteristic temperature dependence of the phonon thermal conductivity peaking near room temperature or slightly above.

In appliance with [32] thermal conductivity, associated only with phonons is given by:

$$\mathbf{\lambda} = 1/3 \; C \times \upsilon \times \Lambda \qquad (1.6),$$

where:

C – the total heat capacity of the phonon gas;

υ – the rms velocity of the phonon gas;

Λ – the mean free path of the phonons.

Heat capacity associated with the phonons in accordance with Debye law is defined as:

$$C = \begin{cases} a\,T^3 & (T < \theta_D), \\ 3Nk_B & (T \geq \theta_D). \end{cases} \qquad (1.7)$$

Therefore, the phonon thermal conductivity at low temperatures increases proportionally $T^3$, and it is associated with an increase of the phonon numbers. When the heat capacity reaches its limit at high temperatures, the thermal conductivity decrease will be associated only with the change of the mean free path of the phonons. Although on the thermal conductivity λ will be influence the collision of phonons with irregularities and the boundaries of the graphite crystallites, nevertheless the thermal conductivity dependence at high temperatures will be wholly connected with the phonon–phonon interaction, or so named anharmonism [33]. The crystal lattice dynamics with account the phonon – phonon interaction is very complicated, but the end result is the phonon mean free path is inversely proportional to absolute temperature:

$$\Lambda \sim T^{-1} \qquad (1.8)$$

Consequently, the phonon thermal conductivity at high temperatures is inversely proportional to temperature. In the graphite at high temperatures in the thermal conductivity are also participating carriers, especial for graphite with a low degree of ordering of the lattice [31].

### 1.1.5. Technological aspects of producing high–strength artificial graphite

Physical and mechanical properties of artificial graphite defined by features of the crystal structure on the both micro and the macro level. Consequently, this structure depends on the nature and quality of raw materials and manufacturing technology features [34]. Graphite is not a self–baking material because it has low values of the self–diffusion coefficient even at treatment temperatures 2000-3000°C. The basis of modern classification for artificial graphite is the grain size.

Classical technological scheme producing artificial graphite includes process steps of preparing the filler and binder from the raw materials, blending composition molding of preforms, their firing and graphitization [35, 36]. The grain size of graphite is usually determined by the size of the filler particles. As a filler for the production of artificial graphite used various types of coke, the structure of which may be quite vary. [34]. The highest value to the production is represented the low–ash cokes, with an ash content not more 1%. At low–temperature carbonization the pitch–shaped mass of oil residues formed the main structural features of coke. Heating of these residues gives the spherical particles. These particles are called *spherules* and they are similar to the liquid crystals. Transformation start temperature is 400 up to 520°C and depending on the type of the carbonation material [37]. With increasing temperature *spherules* getting to interact with each other and as a result of coalescence is appears so named "mosaic". The mosaic structure and its stratification are stored forming a rigid coke frame, which is accompanied by the appearance of a discontinuities and pores.

Low–ash coke are of two types – oil and pitch .The first is obtained by coking of the petroleum residues, the second is when coal–tar pitch transformed into the coke. Properties of petroleum cokes depend mainly on the type of oil residues from which they are obtained. Processing conditions affect to the properties of petroleum cokes not so much. From small oxidized, hydrogen–rich raw materials receive usually easily graphitized materials. It should be noted that the coke particles often have an elongated shape, this anisometric leads to the anisotropy of properties for a final graphite composites.

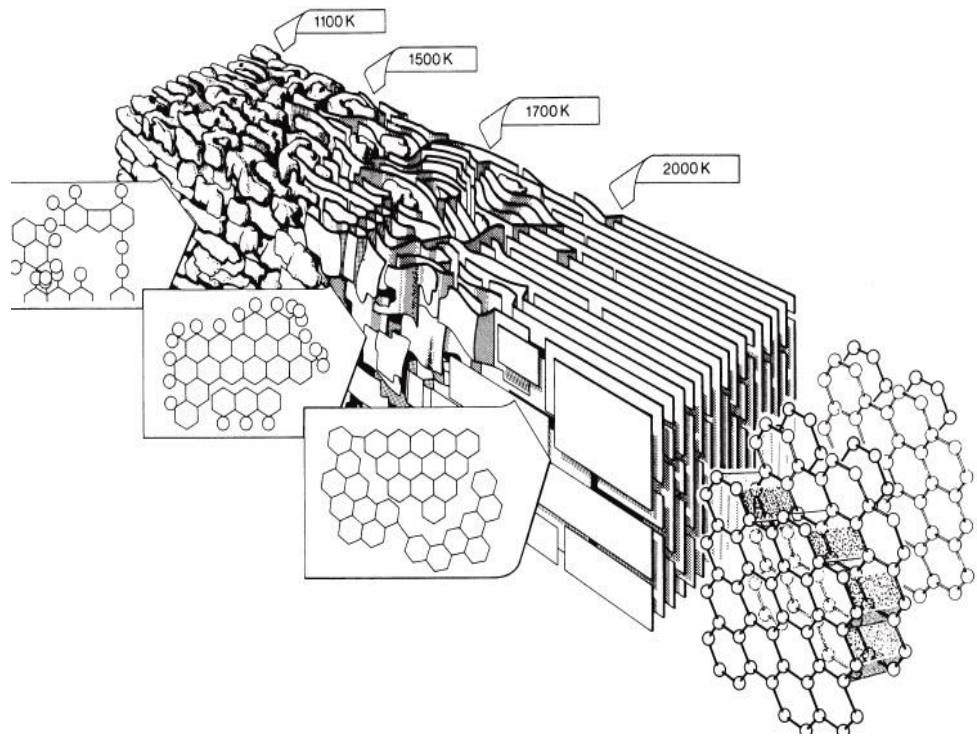

**Fig**. **1.19**. Graphite mesostructure change during heat treatment [38].

Since the end of the XIX century as coal–tar pitch is used as binder. Coal pitch is the residue after distillation of coal–tar into fractions, its hardness can be varied as so the softening temperature. In the technological process of artificial graphite preparation can be distinguish such stages as pre–fragmentation the carbon raw material, burning, drafting the charge, mixing, forming and annealing. Thermal conversion process of raw carbon materials produced in special electric furnaces, the graphitization usually ends at 2400–2800°C.

The process of transformation of the starting amorphous carbon material in highly ordered artificial graphite schematically shows *fig.* 1.9. The structure of the carbon material starts to rebuild at a temperature above 1600 – 1700°C; basal planes are ordered, and the interlayer spacing decreases slightly.

Due to the destruction of the lateral free radicals increases the number of carbon atoms. The formation of three–dimensionally ordered structure of the crystallites occurs at temperatures above 2000, it is accompanied by a sharp increase in their height $L_c$ and diameter $L_a$. The differences of the structural characteristics allow to explain sometimes the difference in the strength of graphite composites. These structural characteristics are, first of all, the density and meso – and microporous material, texturing, anisotropy, as well as the crystallite size and degree of perfection of the crystal lattice.

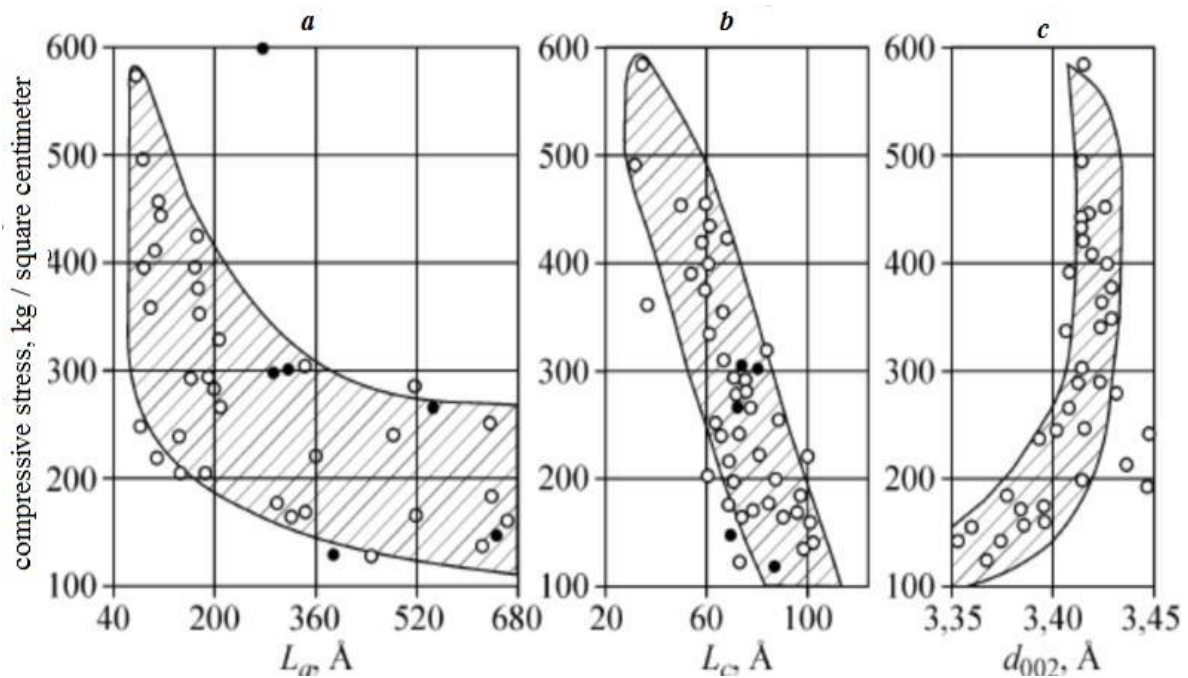

**Fig**. **1.10**. The dependence of the graphite compression strength via to the crystallographic parameters [39].

a, b) – via the crystallite size *La, Lc* c)– interlayer distance $d_{002}$; open circles – experimental materials; dark circles – industrial materials.

The empirical relationship can be obtained when considering the three structural characteristics: crystallite size along the **a** and **c** axes (*La* and *Lc*) and the interlayer spacing $d_{002}$. In particularly, by [39] it looks like:

$$\sigma_{compr.} = const/\, L_a, \qquad (1.9)$$

However, this dependence is typical only for graphite materials not passed of the graphitization process up to the end. Relations between the compressive strength and crystallographic parameters is shown in *fig.* 1.10, here are the data from [39].

## 1.2. Key notions of the destruction dynamics

1.2.1. Structurally–analytical theory of strength is one of the subdivisions of nonlinear dynamics, which is used in all modern concepts of materials science [40, 41]. Among the theories of nonlinear dynamics, in addition to the structural–analytical theory of strength should be called also the physical mesomechanics, the concept of an elementary shift, and fractal mechanics of materials. The structural–analytical theory of strength [41] attempted unification of major achievements of physics and mechanics of destruction with the aim of constructing such equations that would properly consider the physical aspect of phenomena and at the same time allows to produce of engineering calculations. The theory lay on the analytical relations, which proceed from the fact that the driving force in a structured continuum is the effective stress field $\sigma^{*}_{ik}$. The latest determined through the application $\sigma_{ik}$, oriented $\rho_{ik}$ and non–oriented $\lambda_{ik}$ tension:

$$\sigma^{*}_{ik} = \sigma_{ik} - \rho_{ik} + \lambda_{ik} \qquad (1.10),$$

where indices *i, k* generally describe isolated meso–volumes of anisotropic medium in whole. This is the basic equation of the theory, written in the tensor form for micro–level, and it is similar for macro–level. It should be say that in this case is chosen two–level model of the process: applied stress generate micro tension that cause the physical acts of mass transfer and micro–flow passing in the macroscopic deformation. The principal for the theory is the concept of a representative volume, the elementary act of deformation and the laws of a deformation behavior. The properties of the representative volume should be expressed through the mean values of variables that characterize it as a continuous and relatively homogeneous medium.

A representative volume can be viewed as a mathematical point in a continuum, whereby the physical aspects of the theory are averaged on the lower microstructura level and mechanical on the upper. In addition to the equation (1.10) in the mathematical model entered the equation to take into account the kinetic coefficients of structural compliance, fluidity, inhomogeneity, relaxation, and etc. All of these coefficients are presented in the form of functional like a tensor of the fourth rank, along with the elastic compliance and thermal expansion coefficients. Calculations and experiments have confirmed today that the structural–analytical theory of strength able to predict the properties of active plasticity, creep and fracture, as well as can be used to describe non–trivial mechanical properties on the various structural levels of solids.

1.2.2. Physical mesomechanics deformable solids is a new direction in the materials science, which is based on the concept of structural levels of deformation [42,43 ].This concept is based essentially on the next following provisions:

– deformable solid body is a multi–level system in which plastic flow develops as a self-consistent evolution of the stability loss at a different scale levels: micro, meso and macro;

– the three–dimensional structural elements are elementary carriers of plastic flow at the meso level, it may be for example grains, grain conglomerates, subgrains, cell dislocation, domains, second–phase particles, pores, etc. The movement of this carriers is characterized by the scheme "shift + turn";.

–this concept is organically interconnected translational and rotary mode of the motion for the three–dimensional structural elements. Rotary mode of the plastic deformation leads to a self–consistent motion the entire hierarchy of structural levels, and determines the appearance of new dissipative structures:

– main regularities of plastic flow at the meso level are associated with the formation of dissipative mesostructures and fragmentation of the deformable material;
– destruction is the final stage of the solid state fragmentation when it occurs at the macroscale level;
– mechanisms of plastic flow, their carriers and appropriate stage of the curve of "stress–deformation" agrees with the law of similarity (the principle of scale invariance).

Self-consistent deformation in the whole volume of the solid state described by the law, according to which the rotor amount of the deformation flows through the whole hierarchy of the structural levels should be zero. The latter is true for the plastic deformation of the crystal without discontinuity loss.

Destruction is the final stage, when the localized translation–rotation vortices reach the size close to the cross section of the sample, and the rotor of the primary slide in the vortex is not compensated by the summary rotor of the accommodative processes.

It should be noted that the introduction of the concept of structural level into physical mesomechanics of deformable solid state allowed to link continuum mechanics and the theory of dislocations. The universality physical mesomechanics is based on the fact that this theory is built on the principle of gauge symmetry [44]. From this viewpoint we can say that Fiz is an analogue of the other gauge theories, such as Maxwell's electrodynamics. A striking example is the analogy between the wave of plastic deformation and an electromagnetic wave carrying the energy field. Electrical breakdown in gases and the solids destruction as the end energy dissipation stage are also an analogy processes.

The basic equations of mesomechanics for shift rate $\nu$ and turns $\omega$ in the displacement field of the crystal is looks really like Maxwell's equations for the electromagnetic field:

$$\text{div } \nu = g^{ij} \eta^{\alpha}_{i} \acute{\eta}_{j\alpha};$$

$$(\text{rot } \nu) = d\omega/dt;$$

$$(\text{rot } \omega)_{\mu} = \frac{1}{C_t^2} \left[ \frac{d\nu}{dt} \right]_{\mu} + g^{ij} \eta^{\alpha}_{i}(D_{j}\mu \, \eta^{\alpha}_{i}); \qquad (1.11)$$

$$\text{div } \omega = 0.$$

Here $= g^{ij} \eta^{\alpha}_{i} \acute{\eta}_{j\alpha}$ is function characterizing the sources of vortices; $C_t$ - the limit speed of propagation of the gauge field in structurally inhomogeneous medium; $+ g^{ij}\eta^{\alpha}_{i}(D_{j}\mu \, \eta^{\alpha}_{i})$; – field flows due to a change of the rapper in the space; indexes $ij$ are numbered of the interacting meso–volumes, and indexes $\alpha$ and $\mu$ are described the dimension of a heterogeneous medium and can take a value of 1, 2, 3 [41]. Currently physical mesomechanics is getting as the basic methodology for the computer modeling and the computer design of the materials with complex internal structure.

1.2.3. Nonlinear dynamics of the strength theory operates with the spontaneously occurring specifically–ordered forms, that called as the dissipative ones [40, 45]. For any medium if there is an interaction component, the equation can be used:

$$\acute{g}(i) = \alpha g(i) + \beta g(i)g(j) + f(t), \qquad (1.12)$$

where: $\acute{g}_{(i)}$ is the rate of the change of the main parameters, such as the generalized coordinates for selected element; $\alpha$, $\beta$ – the meaning of these constants is determined by the specific task, and the control parameter $\beta$ describes interaction between system elements; $f_{(t)}$ – this function characterizes the probability of fluctuations. Dissipative structures are formed under non-equilibrium conditions with supply of the external energy into the material. Negative feedback in the deformable material is determined of the structure at the quasi-equilibrium stage. Whereas positive feedback determines the self-organization of the dissipative structures in the points of system instability, or so called bifurcation points. At the bifurcation points decreases the non-equilibrium degree of the system as result of an action of the positive feedbacks. With time the non-equilibrium degree of the system increases again, and the cycle repeats up to the destruction of the system.

In the analysis of no equilibrium systems is considered, as a rule, the sequence of the stationary no equilibrium states instead the temporal evolution. For example, during the deformation of metals and alloys in the process of evolution of the system is realized the spectrum of the point-of –bifurcations. These points correspond to the leader-defects responsible for the dissipation of energy at different levels of quasi-equilibrium systems. The transition from one non-equilibrium state to another carried out when the system reaches a certain threshold of energy dissipated by the leader-defect [40].

On the basis of this approach one can be divided the construction materials for several classes by the mechanism of energy dissipation and the dominant type of leader-defect. Physical material treats the elementary acts of plastic deformation and the fracture as separate but interdependent events. They operate in solids, that is a statistical system of a large number of interacting atoms and crystal structure defects.

1.2.4. Kinetic concept of strength theory is showed that the destruction of a solids begins since their loading and it is the kinetic, thermal fluctuations process [46]. The concept that was proposed in this paper not only showed the destruction is a a thermal fluctuations process, but also allowed to substantiate the principle possibility of the forecast for the destruction time. The macroscopic kinetics of destruction, depending on the applied stress $T$ and the absolute temperature $T$ is described by equation of durability:

$$\tau_{lf}(\sigma) = \tau_0 \exp[(U_0 - \gamma\sigma)/kT], \qquad (1.13)$$

where: $k$ is the Boltzmann constant; $\tau_0$ – the time is close to the period of the thermal vibrations of the atoms $\sim 10^{-13}$sec; $U_0$ – initial activation energy of the fracture process, reduces by the applied stress $\sigma$; and $\gamma = qVa$, where $Va$ is activation volume in the elementary act of the destruction; $q$ – the coefficient of the local overstress. This coefficient reaches values since 10 up to 100 in real bodies, and characterizes the difference between the experimental value and theoretical strength. The macroscopic deformation also describes by the kinetic relation:

$$d\varepsilon/dt = d\varepsilon_0(\sigma, T)/dt \exp[(Q_0 - \alpha\sigma)/kT]. \qquad (1.14)$$

There $d\varepsilon/dt$ is the rate of deformation, and the value of $Q_0$ is the energy of activation of the creep. The relation (1.13) and (1.14) have a similar form, but are written for various processes, and generally $Q_0 \neq U_0$ and $\alpha \neq \gamma$.

If the elementary act of the plastic deformation involves a temporary weakening or rupture of the interatomic bonds, the destruction caused by the formation of a voids within the action of an atomic forces can be neglected. The sight of equation (1.13), and the extensive experimental data are evidence that an atomic transformations leading to the destruction carried out by the thermal fluctuations. Thus, the destruction of the macroscopic body is a process of the successive elementary acts of the atomic bonds breaking by the fluctuations of a thermal energy of atoms.

### 1.3. The two-stage model of the solid fracture

The macroscopic time before destruction of the material under load $\tau_{lf}(\sigma)$ can be represented as a two-step process [47, 48]:

$$\tau_{lf}(\sigma) = \tau_1 + \tau_2, \qquad (1.15)$$

where $\tau_1$ – the accumulation time of microcracks and formation of stable macro-cracks in the ensemble of microcracks; $\tau_2$ – the time for the growth of the crack up to the critical dimension, wenn the stability is lost, and $\tau_1 >> \tau_2$

The times $\tau_1$ and $\tau_2$ can be expressed through the average time $\tau_m$ of microcracks formation:

$$\tau_1 = k_1 \tau_m \qquad (1.16)$$

$$\tau_2 = k_2 \tau_m \qquad (1.17),$$

where the formation of microcracks according to [46] is:

$$\tau_m = \tau_0 \times \exp[(U(\sigma)/kT]. \qquad (1.18).$$

In this case is assumed that $k_1$ and $k_2$ are well known structural parameters, and assumed too that for fine-grained and nanostructured materials a conditions for the formation of micro–cracks in the volume and grain boundaries are comparable. This leads to the significant depending of the fine material durability from the grain size [47]. In this case, the material can be imagined as a two–phase, one phase of which is the boundary of a grain, and the other is the volume of grain. In the general case, the probability of microcracks formation at the grain boundaries and in the bulk material can be considered as independent processes. In case if the one of the probability is considerably greater than other, we again come to the formula (1.13).

It is interesting to note that the rate of microcracks accumulation $dC/dt$ is described by equation that looks like (1.13) :

$$dC/dt = C_0 \exp[(U_0 - \gamma\sigma)/kT] \qquad (1.19).$$

This indicates that namely the kinetics of microcracks accumulation eventually determines the durability of the loaded material. According to [48], the predestroying state prediction should be carry out in the control on the process of micro–cracks accumulation by the one or another method in order to prevent the transition to the second stage of the destruction process. The acoustic emission (AE) method [48] could be used to the control, for example.

Fig. 1.11 is shown that the rate of the microcracks accumulation first increases and then becomes flat. When the destruction domain are appeared the rate again rapidly increases up to the fracture. The idea of the two–stage model of destruction [49,50] arose from necessary to analyze the large ensembles of a microcracks for understanding the transition from micro– to macrofracture and use the statistical regularities for quantitative analysis. This concept is illustrated in fig. 1.12.

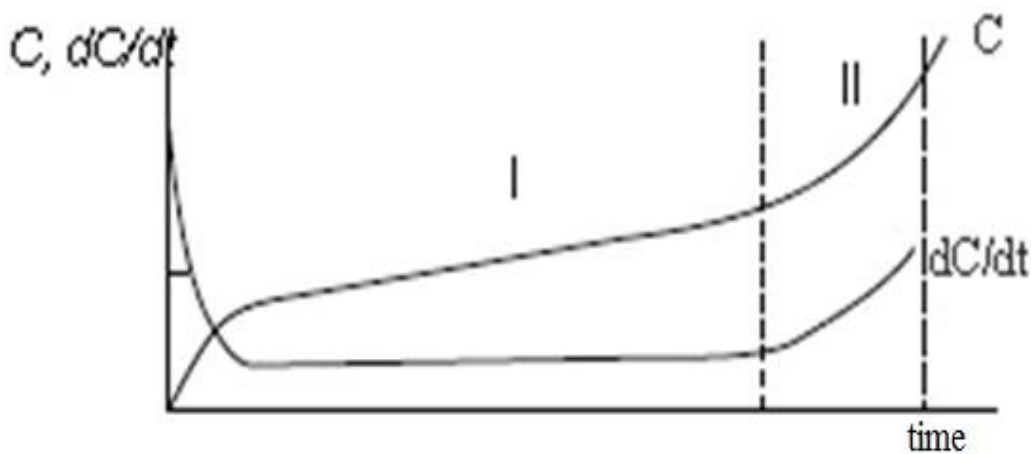

**Fig**. **1.11**. Accumulation of microcracks (C) and the rate of its accumulation (dC/dt) [48]

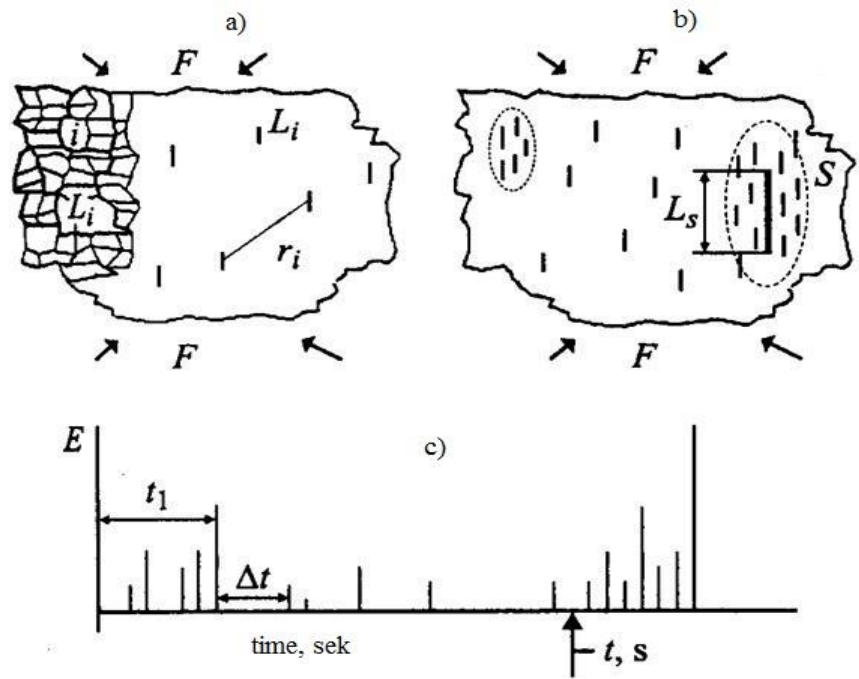

**Fig**. **1.12**. Two stages of destruction: a stationary (a) and focal (b); the acoustic diagram in the formation of microcracks (c); $E$ — energy of the emission, $t$ — time, $F$ is mechanical stress, $Li$ — the size of second rank of initial cracks, $Ls$ — size of second rank, $S$ — area of the destruction [48].

In the first stage (1.12, a) there is an accumulation of single stable microcracks in volume, that is due to fluctuations leads to the formation of ensembles of the closely spaced micro–cracks. These micro cracks are capable to the interacting and fusion, so at the end is formed the fracture domain. The second stage (fig. 1.12, b) is determined by the development of the destruction domain with the appearance of the main crack and the sample is destroyed.

Quantitatively, the transition from the first stage of the cracks accumulation to the appearance to the cracks ensembles or clusters was analyzed in [48]. The concentration parameter is a determining parameter of this transition:

$$K = r_i/L, \qquad (1.20)$$

where $L$ is the cracks size, $r_i$ – the average distance between cracks. Active clustering begins when $K \sim 3$. Equation (1.13) allows to estimate the time before the destruction of the loaded body with logarithmic accuracy, if we know its parameters.

Two–stage models, in turn, allows to formulate the physical principles to predict the macroscopic fracture and significantly improve the accuracy of an estimation of the destruction moment. The prediction *predestroying* state is possible if there is a way to control the damage accumulation and notice the transition from the first stage of the damage into the second one.

In addition, the accumulation of damage is the stochastic process, that allows the use of a number of the statistical parameters to describe the process of damage accumulation. These parameters should be changed regularly during the transition from the quasi–static phase of a damage accumulation into the stage of development of the destruction domain. Thus, the microcracks accumulation increases rapidly at first and then it becomes flat according to the curve *C* of the fig. 1.11.

The intensity of the defects emergence increases again up to the final destruction at the appearance of the hearth destruction domain. The average distance between defects in their chaotic emergence in the first stage remain constant, and sharply reduced when the localized hearth domain is appeared. Especially informative is the parameter associated with the time intervals between chronologically successive acts of the defects appearance. Firstly, it is related to the rate of a nucleation of defects and the curve may describe its accumulation kinetic. The stationary portion where rate of accumulation of defects is constant, this parameter also remains constant and it is reduced by the active development of the fracture domain (fig.1.12, b).

A two–stage model of fracture typical for a very wide range of materials, including metals, polymers, various composites, etc. The structure of the material, homogeneous or heterogeneous, forming a local over–stress is determines the initial size of the microcracks.

Analysis of the fracture characteristics of homogeneous and heterogeneous porous systems [51] indicates that in the "heterogeneous" destruction total applied strain is spread in the large number of the vertices of the microcracks tops. However, at the "homogeneous" destruction strain is concentrated on a few vertices only. As a result the heterogeneous systems can withstand more higher ultimate mechanical stress, i.e. have more high durability.

The studies carried out on various materials: polymers, metals and composites have shown [48, 51] that the size of the initial microcracks is determined by the structure of the material. The latest is not only forms the local over–strain, but also restricts its growth on the borders of heterogeneity.

## 1.4. The properties of the graphite structure change under the irradiation

### 1.4.1. The graphite layings operability in the uranium–graphite reactors

The graphite behavior features in the graphite laying of a reactors are investigated in a number of papers, in particular [19, 37, 52–57]. Graphite, and fine–grained graphite in particular, intended for nuclear reactors, has a more high strength at a high temperatures. This property explains usually by the "healing" of the microcracks that resulting from the self–diffusion of atoms of the carbon at the elevated temperatures [19, 31]. The noticeable increase in the strength of a graphite composites going up to $2400 \div 2500^{\circ}C$, then the strength begins to drop sharply. Such unique character of the strength properties of the graphite composites at elevated temperatures allows to keep operate capacity of graphite laying of the uranium–graphite reactors under the neutron irradiation for many decades [57].

The destruction process dynamics of the destruction of a graphite under high temperatures and neutron irradiation going, as one can be supposed, in two stages, too. This can be detected, for example, by the changes of the electrical conductivity that is associated with the conductivity and tensile strength of a graphite composite under the neutron fluence (fig.1.13).

In [53–57] as a criterion of the operate capacity of the uranium–graphite laying is selected the ratio of the compressive strength and the bending tests as so this ratio does not change during the oxidation or irradiation by the doses that are below of the critical neutron fluence. This empirically founded indicator is very sensitive to the radiation swelling of a samples and begins to grow very rapidly under the supercritical neutron fluence.

Thanks to the works of prof. Yu.S.Virgilev and other it was established the relationship between the relative changes size graphite samples of reactor and the physical properties of a graphite, such as the resistivity, the Young's modulus, the tensile strength etc. as a function of the neutron fluence [3].

To calculate the deformations of the graphite laying during the exploitation period in [58] was proposed a set of the mathematical models and programs with account of the anisotropic shrinking and swelling of the graphite, as well as taking into account the heterogeneity of the distribution of the temperature and neutron field. This allows to estimate the durability and lifetime of graphite laying of the nuclear reactor.

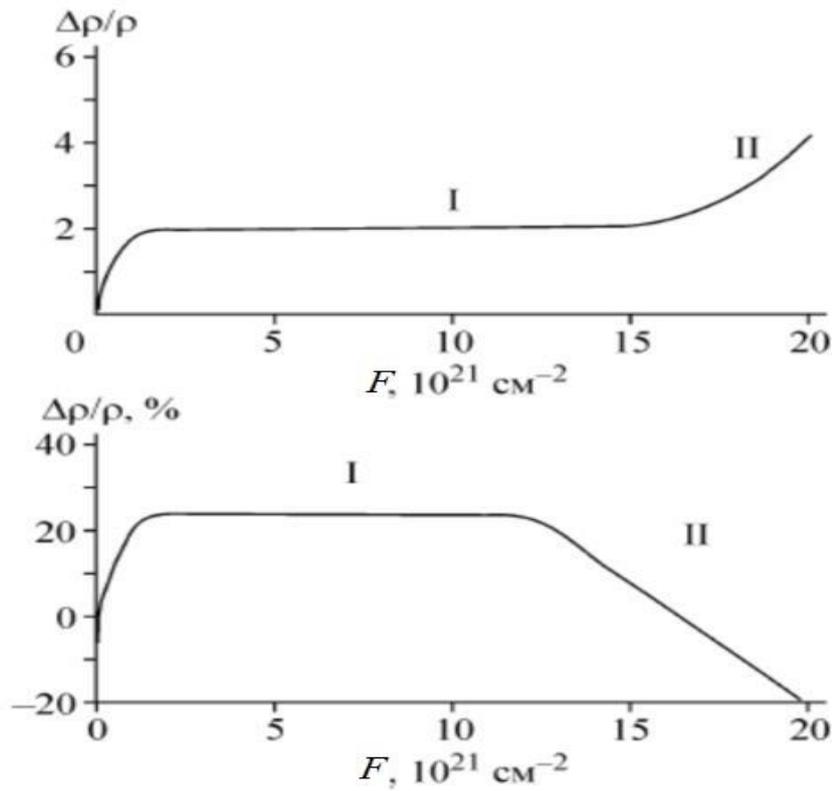

**Fig**. **1.13**. The typical curve of the relative change resistivity depending of the samples of graphite laying from the neutron fluence (above). The characteristic curve of the compressive strength relative change of the samples from the the neutron fluence (bottom) [57].

### 1.4.2. The theory of the graphite radiation–induced forming

The two–stage model of fracture, despite its simplicity and clarity, does not apply as a rule to a complex heterogeneous systems under the combined action of the irradiation, high temperature and mechanical stress. In this situation it is requires a totally different approach, and this approach has been developed in a lot of studies, primarily by *Kelly B.T.* [59], as well as in a number of papers by *Panyukov S.V.* et al. [60–61].

By the [60] «….graphite is viewed as the polycrystal consisting from the ordered crystallite, so one can describe the radiation–induced effects» (fig.1.14). The distant textural order in graphite or significantly weakened or absent. It is depends from the degree of the disorder in the crystallite orientation. The difference in the orientation of neighboring crystallites in the presence of the anisotropy of their properties leads to the appearance of the micro-cracks at the crystallite boundaries. These microcracks are making a decisive contribution to the strength of the structurally disordered solids.

The graphite morphology [61] is defined by two main components: the filler and the binder in the presence of the microcracks ensemble and the technological pores. The binder has a homogeneous fine-grained structure, while the filler structure is hierarchical and the microcrystallite has a more or less perfect crystalline structure. Micro-crystallites form a grains with the different degrees of the texture.

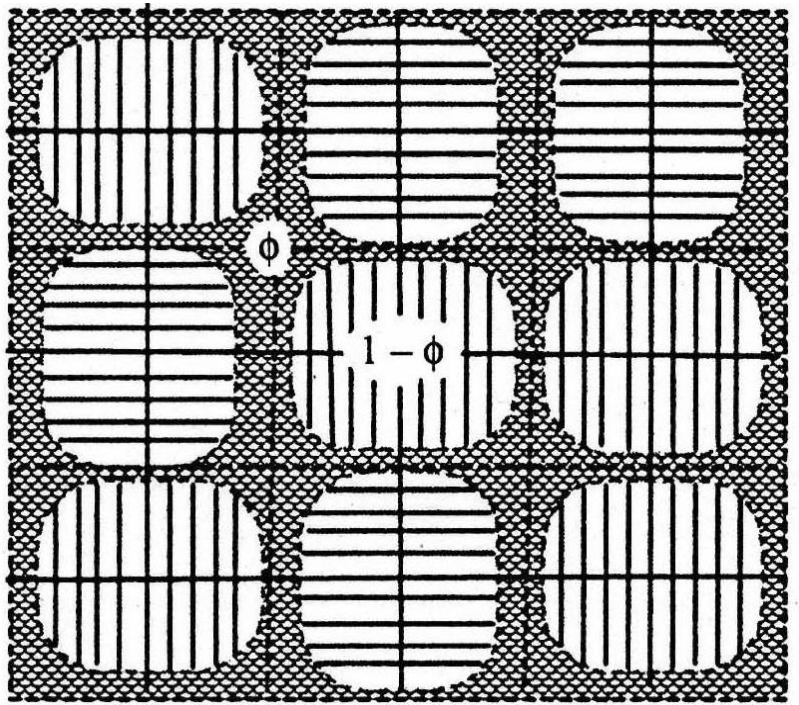

**Fig**. **1.14**. The graphite lattice model with crystallites (*1–φ*). The homogeneous binder fills the matrix nodes. Anisotropy axis is directed along the axes of the lattice [60].

The microcracks ensemble is a consequence of the initially strong anisotropy of the properties at the microcrystallite level, while the anisotropy of considerably weakened at the level of the grain. The microcracks ensemble is evolving under the irradiation, and it is the greatly reason for the observed radiation effects. At the polycrysta level on the whole the irradiation effects are both isotropic or orthotropic, depending on the method of the graphite producing.

The driving force behind all of these effects is the radiation induced shape change of the microcrystallite. The microstructure changing by the irradiation is occurs mainly in the basal plane of the microcrystallite. The theory allow to describe radiation effects in the graphite at the macro level currently missing, although these effects are described empirically and understood at a qualitative level.

In [61] was proposed an analytical model, which managed to link the development of the internal stresses with the evolution of an ensemble of microcracks on the basis of the consideration of an orthotropic continuum with the randomly distributed crystallites. The model takes into account two main effects that determine the elastic properties of graphite and not taken into account until now. This is the internal stress increase of the graphite due to the deformation of the crystallites under irradiation and changes in the elastic properties of graphite due to the evolution of a microcracks ensemble.

The lattice model of the polycrystal is considered to describe the elastic properties of the graphite (fig. 1.14). In this model, the crystallites are in the lattice sites with coordinates $x_j$. The model assumes the presence of a disorder in the orientation and size of the crystallites and its deformation is occurs under the action of the random forces. In the hypothetical case of a regular lattice of the crystallites, these forces are also regular and correspond to the compression along the major axis of the crystal and the more weaker stretching along the other axes.

The microcrystallite size distribution is substantially bimodal in the real disordered graphite, because along with large crystallites of the filler has a lot variety of the small crystallites from binder. If the more large crystallites can be accounted in the lattice model clearly, the small ones from the binder are hindered for a such analysis by a number of reasons. It is also assumed in this model that the elasticity of the structurally disordered solids is substantially determined by the presence of the micro-cracks between all neighboring crystallites. When deformation of graphite as a result of the evolution of internal stress and under the influence of external forces is a change of form and volume of microcracks. As a result of the evolution of the internal stress, and under the influence of external forces is a change the shape and volume of the micro-cracks at the deformation of graphite.

The model assumes that the crystallites deformation and the internal stresses $p_i$ of the crystallites are proportional to the neutron fluency $F$, so that $p_i = p_0 \, F \, (1 - c_i)$. The amount of the dangling bonds between the crystallites $c_i$ initially decreases with increasing fluence due to a disappearing of the small micro-cracks. At the end it is turns out so that the dependence of the average number of microcracks via the fluence $F$ can be approximated by a linear function.

$$c_i = c_{i0} \left(1 - \beta_i \times F\right) \qquad\qquad (1.21)$$

where $c_{i0}$ is the relative share of the microcracks in the absence of the irradiation.

The values of the parameters $p_0$, $c_{i0}$ and $\beta i$ are depend on the microscopic mechanisms of the microcracks formation and in frame of this model are seen as the initial parameters. These parameters depend on the filler and binder composition, graphitization temperature, etc. Calculations carried out in the framework of the lattice model show good agreement between theory and experiment (fig. 1.15).

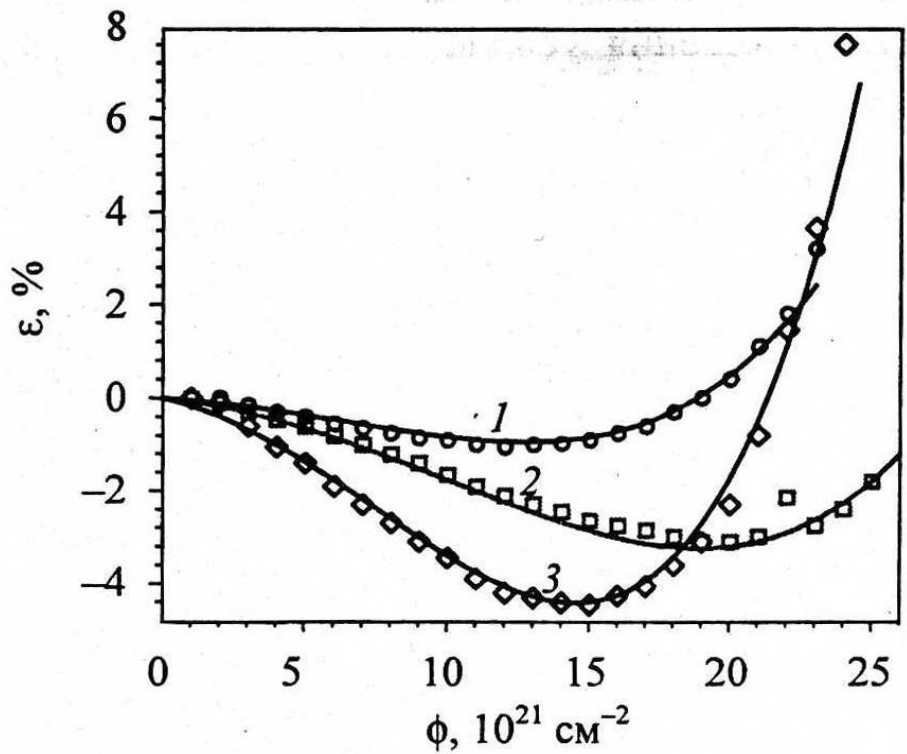

**Fig**.**1.15**. The relative change in the cross (1) and longitudinal (2) size and the volume change (3) from the radiation dose to the graphite GR–280. The calculated curves are shown by the solid lines. [61].

The new ensemble of the microcracks develops in the case of a large irradiation swelling, which should be accounted for the internal radiation stresses finding. The parameter $\beta_i$ in this case is selected from the conditions of which the experimental stress–strain curve of the graphite is repeated and, in fact, is a fitting one.

## Conclusion

Thus, it can be said that the strength and durability of graphite is directly related to the micro – and mesostructure carbon composite material, as well as characteristics such as density and anisotropy, crystallite size, pore size and its nature, i.e. with the material *structure* in the wide sense of the word. The strength change of the material occurs when there are the significant modification of its phase composition, microstructure and/or a porous structure. In other words, when there are the phase transformations processes, sintering, recrystallization, formation or healing of a defects, all of these processes together, or at least one of them.

Therefore, it is so important to control by the physical methods such characteristics as the crystallite size, microdistortions, the ratio of crystalline and amorphous components, size and total pore volume, etc. Based on these data it is not always possible to predict the strength characteristics, but can say with great certainty about the stability of the material structure, and hence about the preservation at least the initial strength.

The foregoing suggests the necessity for the studies of the graphite material structure and defects to forecast the lifetime of its operability on the basis of the most common reasons of structural– analytical theory of strength. Thus, in particular, more than important is the question about the interrelated phenomena of the strength, creep, and the self–diffusion. It should be noted that in the graphite composites, as well as in the metals the role of the intergrain boundary is very significant. Thus, in particular, for metals have been shown [62] that this creep activation energy close to the activation energy of a self–diffusion, under certain conditions may become the controlling mechanism of the energy dissipation.

One can also assume that in the graphite as well as in metals, the main mechanism of dissipation of excess energy supplied due to proton or neutron irradiation is the plastic deformation, and as the dominant leader–defect can be regarded as a different types of the dislocations and vacancies, as so other defects of the crystal structure. The boundary dislocations, interstitials and impurity atoms, the packaging layers violations, and also the features of the graphite mesostructure can play a decisive role in the stability and durability of the graphite materials.